\documentclass[sigconf]{acmart}
\AtBeginDocument{%
  \providecommand\BibTeX{{%
    \normalfont B\kern-0.5em{\scshape i\kern-0.25em b}\kern-0.8em\TeX}}}

\setcopyright{acmcopyright}
\copyrightyear{2022} 
\acmYear{2022} 
\setcopyright{acmcopyright}\acmConference[WSDM '22]{Proceedings of the Fifteenth ACM International Conference on Web Search and Data Mining}{February 21--25, 2022}{Tempe, AZ, USA}
\acmBooktitle{Proceedings of the Fifteenth ACM International Conference on Web Search and Data Mining (WSDM '22), February 21--25, 2022, Tempe, AZ, USA}
\acmPrice{15.00}
\acmDOI{10.1145/3488560.3498478}
\acmISBN{978-1-4503-9132-0/22/02}

\usepackage{balance}
\usepackage{breakurl}

\begin{document}

\fancyhead{}

\title{Modeling Users' Contextualized Page-wise Feedback \\ for Click-Through Rate Prediction in E-commerce Search}



\author{Zhifang Fan$^{1}$, Dan Ou$^{1}$, Yulong Gu$^{1}$, Bairan Fu$^{2}$, Xiang Li$^{1}$, Wentian Bao$^{1}$, Xin-Yu Dai$^{2}$, \\ Xiaoyi Zeng$^{1}$, Tao Zhuang$^{1}$, Qingwen Liu$^{1}$}

\affiliation{
  \institution{$^{1}$Alibaba Group, China \quad $^{2}$Nanjing University, China \\
  }
  \country{}
}
\email{
{shixin.fzf,oudan.od,leo.lx,wentian.bwt,yuanhan,zhuangtao.zt,xiangsheng.lqw}@alibaba-inc.com}
\email{guyulongcs@gmail.com, fubairan@smail.nju.edu.cn, daixinyu@nju.edu.cn}

\renewcommand{\shortauthors}{Fan and Ou, et al.}

\begin{abstract}
Modeling user's historical feedback is essential for Click-Through Rate Prediction in personalized search and recommendation. Existing methods usually only model users’ positive feedback information such as click sequences which neglects the context information of the feedback. In this paper, we propose a new perspective for context-aware users' behavior modeling by including the whole page-wisely exposed products and the corresponding feedback as contextualized page-wise feedback sequence. The intra-page context information and inter-page interest evolution can be captured to learn more specific user preference. We design a novel neural ranking model RACP(i.e., Recurrent Attention over Contextualized Page sequence), which utilizes page-context aware attention to model the intra-page context. A recurrent attention process is used to model the cross-page interest convergence evolution as denoising the interest in the previous pages. Experiments on public and real-world industrial datasets verify our model's effectiveness. 
\end{abstract}




\maketitle

\section{Introduction}
\begin{figure}[htbp]
    \centering
    \includegraphics[width=0.5\textwidth]{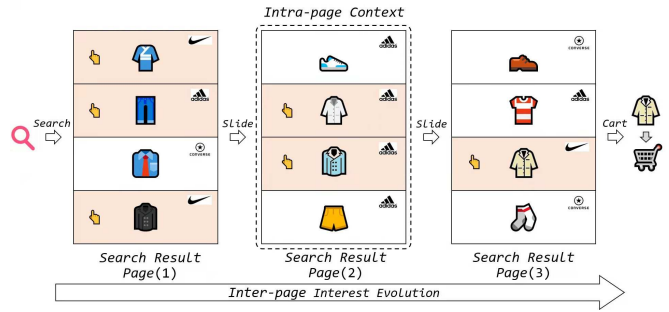}
    \caption{The illustration of a user's feedback in the sequence of Search Result Pages~(SRPs). Intra-page context: the user's behavior (click or not) on a item may be influenced by other items in the same SRP. Inter-page Interest evolution: the user's behaviors in different SRPs reflect the evolution of her interest.}
    \label{fig:motivation}
\end{figure}
Search Engines and Recommender Systems are playing significant roles in E-commerce companies (e.g., Amazon, Alibaba)~\cite{carmel2020multi,ni2018perceive,zhou2018deep}. 
In a typical e-commerce search engine, as shown in Figure~\ref{fig:motivation}, after a user enters a query, the ranking system in the search engine will select a list of items and demonstrate them in a Search Result Page (SRP) to the user.
After the user reaches the end of the current page or enters a new query, the search engine will generate a new SRP.
The user usually browses many SRPs and interacts with items before she purchase a product.
Modeling the users' feedback in these sequential SRPs is significant for ranking in the search engine.

Click-through rate (CTR) prediction, which is one of the critical task for ranking in Search Engines and Recommender Systems, has received much attention from both the research and industry community~\cite{cheng2016wide, he2017neural, he2014practical, shan2016deep, zhou2018deep,liu2020decoupled}. 
Deep learning based methods ~\cite{covington2016deep, cheng2016wide, zhou2018deep, zhou2019deep, quadrana2017personalizing, gu2020hierarchical, gu2020deep} have achieved state-of-the-art performance in CTR prediction by modeling users' sequential behaviors.
They usually model users' positive feedback information (e.g., click or purchase behaviors) using sequential models.
However, these methods neglect users' negative feedback (e.g., skip the exposed item and not click on it).
Recently, some pioneering studies (DFN~\cite{xie2020deep}, DSTN ~\cite{ouyang2019deep}) have demonstrated the effectiveness of modeling both users' positive and negative feedback for CTR prediction.
However, they treat users' positive and negative feedback separately, and represent users' feedback as a clicked item sequence and a non-clicked item sequence, which cannot generate the mutual context between clicks and non-clicks and ignores other page context information in the page-sequence.
It may lead to inferior performance in CTR prediction.

To solve these problems, in this paper, we proposed the idea of extracting the users' interest from their Contextualized Page-wise Feedback. We generate the user behavior sequence as the whole part of exposed items and the corresponding interactions including both the click and slide without a click.

Then we aim to extract the user's interest from two aspects:
\begin{itemize}
\item \textbf{Intra-page context information}. 
User's feedback are influenced by the surrounding items and the whole page context and different types of feedbacks are also contextual to each other(click, no click, add to cart, purchase, etc.).
As shown in Figure \ref{fig:motivation},
Modeling the Intra-page context information is important for preventing over-sensitivity in only modeling the positive feedback and better inferencing user's interest: 
(1) The positive feedback maybe noisy. 
For example, a click behavior on a item which has a specific brand may not represent that the user is really interested in this brand when most of the items on the pages belong to this brand. 
A price-sensitive user’s click on the cheapest items if all the exposed items in one page are in high price range doesn’t reflect the preference on this price range and the user’s interest can only be learned by including the intra-page context.
(2) The user's behaviors on an item are usually influenced by other items.
The users often like to compare items in the same SRP.
If the products in the SRP have multiple different brands and a user click only one of them, it may mean that the user is especially interested in that brand. 

\item \textbf{Inter-page interest evolution}. As shown in Figure \ref{fig:motivation}, the user interests in the previous page may be divergent and uncertain, but as the user continues to browse the pages, her interest in the next pages may become more apparent and convergent.
The user’s intent and interest when search can be considered as a gradual convergence process. The user’s dynamic intent in the previous pages can be denoised by the interactions in the following pages since the following interactions is more relevant to the final decision. Some methods also propose to modeling the interest evolution (DIEN~\cite{zhou2019deep}, DSIN ~\cite{DBLP:conf/ijcai/FengLSWSZY19} and HGRU ~\cite{quadrana2017personalizing}), but they only consider the positive feedback which neglect the context info and lack of the complete page-wise interest evolution.
\end{itemize}

To achieve this goal, we proposed a novel CTR prediction model called \textbf{RACP} (i.e., Recurrent Attention over Contextualized Page sequence), which adopts a hierarchical architecture to model user's interest based on the user's Contextualized  Page-wise Feedback.
Firstly, The Intra-page Context-aware Interest Layer is used to capture the user's interest in each SRP, where each item's attention weights will be influenced by the page context feature and surrounding items and surrounding feedbacks.
Secondly, The Inter-page Interest Backtracking Layer aims to extract the users' interest from previous pages.
A backward gated recurrent units network is used to update the intent query vector for previous pages based on current page's aggregated interest recurrently. 
Finally, the Page-level Interest Aggregation Layer generates the user's whole interest representation across multiple SRPs.
The final representation of users' feedback will be feed to a deep neural network to predict the click-through-rate probability. The open sourced implementations are released at \url{https://github.com/racp-submission/racp}.

\vfill\eject

Our major contributions can be summarized as follows:
\begin{itemize}
\item We are the first to study the user’s page contextualized behavior sequence with the whole page-context for CTR prediction. The intra-page context can be incorporated for better learning of the user’s complete interest in one page. To the best of our knowledge, there is no work studying the structural contextualized page-sequence modeling for personalized ranking and CTR prediction.

\item We devise a novel method RACP that effectively models the intra-page context and the page-sequence structure to learn user's dynamic interest. The model can model the user's interests on each page based on the user's future intent and the page context information.

\item Extensive experiments on both public dataset and real-world production dataset demonstrate the effectiveness of our model.  We also conduct ablation study to verify the effectiveness of the components in our method. We also successfully deploy the RACP model in the real-world large-scale e-commerce search engine in Taobao Search.
\end{itemize}

\section{Related work}
\subsection{CTR prediction} 

Click-through rate~(CTR) prediction, which aims to estimate the probability of a user clicking on the item, is one of the core tasks for ranking in industrial applications, such as Search Engines~\cite{ni2018perceive,carmel2020multi}, Recommender Systems~\cite{covington2016deep, cheng2016wide, feng2019deep,xie2020deep}, Online Advertising~\cite{zhou2018deep, zhou2019deep,ouyang2019deep} and so on.

Deep learning based methods  ~\cite{covington2016deep, cheng2016wide, zhou2018deep, zhou2019deep, feng2019deep, quadrana2017personalizing, gu2020hierarchical, gu2020deep, gu2021self} have achieved state-of-the-art performance in CTR prediction by modeling users' sequential behaviors.
They usually focus on modeling users' positive feedback~(e.g., click or purchase behaviors) to capture the users' preference ~\cite{hidasi2015session, hidasi2018recurrent, sun2019bert4rec, quadrana2017personalizing, lv2019sdm, kang2018self, liu2018stamp}.
Some methods ~\cite{hidasi2015session, sun2019bert4rec, quadrana2017personalizing} utilize Recurrent Neural Networks(RNN) to capture the sequence dependency among the user's behaviors.
Some research work ~\cite{lv2019sdm, kang2018self, liu2018stamp} incorporate attention mechanisms to model the adaptive user preference concerning the target item.

Recently, some pioneering work (DFN~\cite{xie2020deep}, DSTN ~\cite{ouyang2019deep}) highlight the importance of modeling both users' positive and negative feedback for CTR prediction.
Besides the clicked behaviors, there are abundant exposure data in users' historical behaviors called unclicked behaviors (i.e., items which are impressed to the user but are not clicked). These unclicked behaviors also help model the users' preference~\cite{ouyang2019deep, lv2020unclicked, xie2020deep}. 
~\cite{xie2020deep} view click behaviors as strong feedback to guide the positive preference extraction from unclicked behavior sequences. 
\cite{ouyang2019deep} considers the clicked and unclicked behaviors as heterogeneous auxiliary data to help the user preference modeling. 
~\cite{lv2020unclicked} designed a confidence fusion network to gain better user representation. 
Although these works model the user preference with users' clicked or unclicked behaviors, they did not consider the page context information in users' behaviors. 

Session-based recommendation ~\cite{hidasi2015session, li2017neural, quadrana2017personalizing, feng2019deep, lv2019sdm} also model the user interests among their historical sessions. 
However, they are different from our work because they focus on the temporal information and only consider the positive feedback.
We investigate the spatial page context information of user's behavior and capture both positive and negative behaviors.

\begin{figure*}[htbp]
    \centering
    \includegraphics[width=0.88\textwidth]{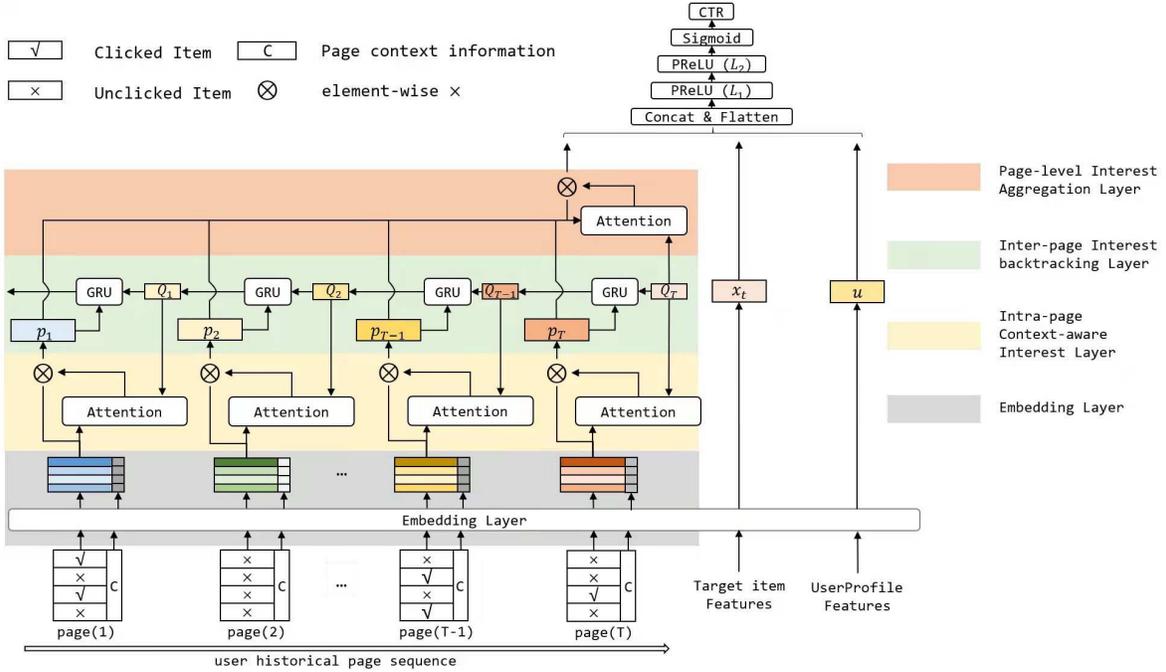}
    \caption{The architecture of RACP.}
    \label{fig:RACP}
\end{figure*}

\subsection{Page Context Modeling} \label{label:reated_pagecontext}
The most related work about page context modeling lie in the field of  re-ranking~\cite{zhuang2018globally,ai2018learning,pei2019personalized}. Re-ranking is a follow-up stage following the ranking model.
They focus on refining the relative order based on the output of the ranking model. ~\cite{zhuang2018globally} and ~\cite{ai2018learning} are RNN-based methods for re-ranking. ~\cite{pei2019personalized} proposed using the self-attention based transformer model to model the influence between items in the list directly. 

However, these work are totally different from ours, and their task is re-ranking by modeling the mutual influence inside the candidate list, while our method is used for ranking by only modeling users' feedback in historical pages information.
DeepPage~\cite{zhao2018deep} is a reinforcement learning framework that focuses on generating a page of items in recommendation systems.
They are significantly different with our CTR prediction task.

\section{Problem Formulation}
\textbf{The Click-Through Rate Prediction problem.}
The click-through rate prediction task for a search ranking system is typically formulated as a binary classification task. 
The training samples are collected from history logs, where the labels are users' feedback information (i.e., click or not).

\section{Method}

The overall architecture of RACP is illustrated in Figure ~\ref{fig:RACP}. RACP can be separated into four parts from bottom to top: Embedding Layer, Intra-page Context-aware Interest Layer, Inter-page Interest Backtracking Layer, and Page-level Interest Aggregation Layer.

\subsection{Input and Embedding Layer}

The Embedding Layer is used to represent the following input of the model into low dimensional vectors. For each item (e.g., query), as previous did ~\cite{covington2016deep}, we use the embedding layer to transform the high dimensional sparse ids of the attributes (e.g.\ query id, word segments) into low dimensional dense representations, and then concatenates the vectors of the item's attributes into a single embedding vector.The embedding tables are initialized as random numbers and jointly learnt with our model. 

\begin{itemize}
\item \textbf{Item profile}.  Item profile contains product id, category id, brand id, shop id and  statistical features (e.g., numbers of click and purchase).
The target item embedding is denoted as $x_t \in \mathrm{R}^{D_t}$.
\item \textbf{User profile and Query profile}. User Profile and Query profile is used to modeling the personalization and intention information. User Profile contains user’s characteristics, such as age, gender, purchase power and so on. Query profile consists of query id, query string, word segments and predicted product category ids. 
The current query embedding is denoted as $q_{T} \in \mathrm{R}^{D_q}$.
The user embedding is $u \in \mathrm{R}^{D_u}$.
\item \textbf{User's historical Page-wise Feedback}. User's Contextualized Page-wise Feedback contains users' positive and negative feedback in the sequence of historical Search Result Pages~(SRPs).
The sequence of historical SRPs is denoted as $\{p_i \}^{T}_{i=1}$,  where $T$ is the number of historical pages.
For each historical SRP $p_i$, the list of items in the page are represented as $\{x_{i,j} \in \mathrm{R}^{D_x}\}^{L_a}_{i, j=1}$, where $L_a$ represents the number of items in the page. We represent the context features (e.g., query in the SRP) in SRP $p_i$ as $c_i \in \mathrm{R}^{D_c}$.
$D_q, D_t, D_u, D_x, D_c$ are the numbers of vector dimension.
The corresponding query in the SRP $p_i$ is $q_i$, which may or may not be the same with current query $q_{T}$. 
We restrict that $q_i$ and $q_{T}$ are in the same product category (e.g., T-shirt) to filter irrelevant search behaviors with current query $q_{T}$.

\end{itemize}

\subsection{Intra-page Context-aware Interest Layer}
The Intra-page Context-aware Interest Layer aims to extract the user's interest from user's feedback and context information in each Search Result Page (SRP). Not every item in the user’s history is equally important for predicting current interest.
Existing methods ~\cite{zhou2018deep,zhou2019deep,ouyang2019deep} typically incorporated an attention mechanism to compute the important weight for the interacted items but deciding which item is more important is context-specific and chal- lenging for vanilla attention over click sequences. To address this challenge, we propose the organize the context as the contextualized page sequence and consider the page-context info when modeling the historical behaviors.

This layer adopts a page-wise attention mechanism. Firstly the historical exposed items are organized by a sequence of pages. 
For each page $p_i = \{x_{i,1}, \cdots, x_{i,j}, \cdots, x_{i,L_a}\}$, it consists of $L_a$ items' embedding vectors and the user's feedback (i.e., click or not) $f_{i,j}$ on these items.
For each item $x_{i,j}$ in the page, we compute page-context feature $c_{i,j}$ based on information in the page:
(1) The searched query $q_i$ and the category of the query in the page.
(2) The number of clicks in the page.
(3) The number of items that has the same brand with the item $x_{i,j}$.
(4) The number of items that has the same seller with the item $x_{i,j}$.
(5) The rank of each item in this SRP according to the price, sales count respectively.
These features will also be mapped into vectors by embedding tables and generate the page-context feature $c_{i,j}$ for each item $x_{i,j}$. 
Then the attention weight $a_{i,j}$ of the item $x_{i,j}$ is computed as follows:
\begin{equation}
\begin{aligned}
    a^*_{i,j} &= W^a_i * [Q_i; x_{i,j}; f_{i,j}; c_{i,j}], \\
    a_{i,j} &= \frac{ exp(a^*_{i,j}) }{ \sum_{j=1}^{L_a} exp(a^*_{i,j}) }.
\end{aligned}
\end{equation}
where $W^a_i$ is the weight matrix, $x_{i,j}$ is the embedding vector of the item, $f_{i,j}$ is the embedding vector of user's feedback on the item, $Q_i$ is the context vector from the following page (the detail is introduced in Section \ref{SubSecionInterPage}). It should be noted that the softmax operation is conducted inside each page. Therefore it can be considered as a comparison process in each page. The weights of the items is not only decided by the item self but also the context information in the page.

Then the representation for each page can be computed as the aggregation of the whole-page info as :
\begin{equation}
    p_i = \sum^{L_a}_{j=1} a_{i,j} * [x_{i,j}; f_{i,j}; c_{i,j}].
\end{equation}
In this way, a page-context aware user's interest vector is generated to capture both feedback and context information for each page. $p_i$ will be fed into the following layer to learn the general representation for the user's interest.

\subsection{Inter-page Interest Backtracking Layer}\label{SubSecionInterPage}
When modeling users' historical actions, two crucial aspects should be considered: long-term global target-relevance and short-term local interest consistency. The typical methods using global target-attention have an advantage in modeling the long-term global target-relevance while ignored the interests consistency among the page sequence. Some historical page's information can be better revealed by its next several pages instead of the target item. 
Therefore, we propose a GRU based Query abduction module to fuse the long-term global target-relevance information and short-term local interest consistency.

Specifically, we can represent the user's  current interest query as $Q_c$ which is a embedding-based representation for the features of target item or the user's feature and input query which represent the user's current intent. Traditional attention use the current request's queryr or target item as $Q_c$ to compute the attention weights for all historical items like:
\begin{equation}
\begin{aligned}
    {{a^*}_{i,j}}^{\text{Global}} &= {W_i}^a*[Q_c; x_{i,j}; f_{i,j}; c_{i,j}], \\
 {a_{i,j}}^{\text{Global}} &= \frac{ exp({{a^*}_{i,j}}^{\text{Global}})}{ \sum_{j=1}^{L_a} exp{({a^*}_{i,j}^{\text{Global}}) }}.
\end{aligned}
\end{equation}
But this global attention method may suffer from the interest gap for computing the intra-page attention in a page long before. 
To solve this problem, we propose to integrate local interest in current page and long-term interest in following pages in a recurrent way. 
For the last page $p_T$, we define $Q_T=Q_c$.
The computing of the attention query $Q_t$ of each previous page $p_t$ (identical to $Q_i$ for $p_i$) is formulated as :
\begin{equation}
\begin{aligned}
    r_t &= \sigma(W_{ir} p_t + b_{ir} + W_{hr} Q_{(t+1)} + b_{hr}), \\
    z_t &= \sigma(W_{iz} p_t + b_{iz} + W_{hz} Q_{(t+1)} + b_{hz}), \\
    n_t &= \tanh(W_{in} p_t + b_{in} + r_t * (W_{hn} Q_{(t+1)}+ b_{hn})), \\
    Q_t &= (1 - z_t) * n_t + z_t * Q{(t+1)}.
\end{aligned}
\end{equation}
where $r_t$, $z_t$, $n_t$ represents the reset, update and new gates respectively,
$p_t$ is the representation of user's interest in current page, $Q_t$ is the attention query vector, 
$\sigma$ is the sigmoid function and $*$ is the element-wise multiplication. This backward gated cell will update the query page by page, and the update gate $z_t$ will control the ratio of the keeping the global interest and the local interest. For example, if a page has no click interaction, the gate may learn to retain more global interest. 

This process also stimulates the user's interest and interaction evolution during user browsing SRPs, since the interest of the previous pages will influence the user's interactions on the next page. We can infer the user's interest through the user's interaction information on the next page, so we call it interest backtracking. In this way, RACP will learn the dynamical page query to capture a more precise representation for each historical page and learnt to denoise the more divergent interest in the previous page. The dynamic query representation will be used to compute each page representation ${p_i}$ depicted in the previous section.


\subsection{Page-level Interest Aggregation Layer}
The Page-level Interest Aggregation Layer aggregates user's interest in multiple SRPs and generates  the  user's final interest vector as $s$ :
\begin{equation}
\begin{aligned}
    \beta^{*}_i &= W_s * [Q_{T};p_i], \\
    \beta_i &= \frac{ exp(\beta^{*}_i) }{ \sum_{i=1}^{T} exp(\beta^{*}_i) }, \\
    s &= \sum^T_{i=1} \beta_{i} * p_i. \\
\end{aligned}
\end{equation}
where $W_s$ is the weight matrix, $p_i$ is user's interest vector in each SRP page, and $Q_{T}$ is concatenation of the embedding vectors of current query, user, and target item. 
This attention module may capture that different pages have different importance for the final prediction. The final user behavior representation $s$ is also a high-level aggregation for all the contextualized page representations. The output dimension of $s$ and the hidden sizes of the attention module and GRUs are all set as $K$.

\subsection{Model learning}
The aggregated user's behavior representation $s$ will be concatenated with the user embedding $u$, current query embedding $q_{T}$ and the target item embedding $x_t$.
The concatenated vector is feed into multilayer perceptron (MLP) to compute the final prediction score:
\begin{equation}
    y = \sigma (\text{MLP}([s; u; q_{T}; x_t])).
\end{equation}
where $\sigma$ is the sigmoid function.

The loss funtion is formulated as:
\begin{equation}
    L_{\theta} = - \sum_{k=1}^N[y_k \log y_k + (1-y_k)log(1-y_k)].
\end{equation}
where the $\theta$ is the trainable parameters and N is the size of the training dataset. 

\begin{table}[tbp]
    \caption{Statistics of the datasets. (``Page Seq Ratio'' represents the ratio of the samples that have previous page sequence)}
    \label{tab:statistics}
    \centering
    \begin{tabular}{c|c|c|cl}
        \toprule
            Dataset & \# Samples &  Page Seq Ratio & Avg \# page length\\
        \midrule
            Avito & 4M & 100\% & 4.2 \\
        \midrule
            Taobao & 53,266M & 78.43\% & 6.7 \\
        \bottomrule
    \end{tabular}
\end{table}

\begin{table*}[htbp]
\caption{Performance of different methods for CTR predictio. ``$*$'' indicates the statistically significant improvements (i.e., p-value < 0.01) over the best baseline.}
\begin{tabular}{l|ll|lll|l}
\toprule
            & \multicolumn{2}{c|}{Avito}                                          & \multicolumn{3}{c|}{Taobao}                                                &                \\ \cline{2-6}
Methods     & AUC                              & RelaImpr                        & AUC                              & PV\_AUC                          & RelaImpr & sequence types      \\ \midrule
NCF~\cite{he2017neural}         & 0.7703                           & 0.00\%                              & 0.7290                           & 0.6418                           & 0.00\%     & No           \\
DeepFM  \cite{guo2017deepfm}     & 0.7714                           & 0.41\%                           & 0.7290                               &                    0.6433             &      0.00\%    & No           \\ \midrule
YoutubeNet~\cite{covington2016deep}  & 0.7846                           & 5.29\%                           & 0.7468                           & 0.6470                               &   7.77\%        & click sequence         \\
DIN~\cite{zhou2018deep} & 0.7849                           & 5.40\%                           & 0.7494                           & 0.6494                           & 8.47\%    & click sequence         \\ \midrule
DFN~\cite{xie2020deep}          & 0.7853                           & 5.55\%                           & 0.7506                                & 0.6467                                &  9.43\%         & click sequence + unclick sequence \\ 
DSTN~\cite{ouyang2019deep}        & 0.7865                           & 5.99\%                           & 0.7535                           & 0.6517                           & 10.70\%    & click sequence + unclick sequence\\ \midrule
\textbf{RACP}         & \textbf{0.7944$^*$} & \textbf{8.92\%$^*$} & \textbf{0.7623$^*$} & \textbf{0.6558$^*$} & \textbf{14.50\%$^*$}    & page-wise sequence          \\ 
\bottomrule
\end{tabular}
\label{main_results}
\end{table*}

\section{Experiments Setup}

\subsection{\textbf{Datasets.}} 
We conduct our research on a public benchmark dataset \textit{Avito}\footnote{https://www.kaggle.com/c/avito-context-ad-clicks/data} and a commercial dataset \textit{TaoBao}. 
The statistics of these two dataset is shown in Table \ref{tab:statistics} and the basic information about the two datasets is summarized as follows: 

\begin{itemize}
\item \textbf{Avito dataset}: Avito dataset is used for the Avito context ad clicks competition, and it contains a random sample of ad logs from avito.ru. Avito dataset contains the user search information and item metadata, include the list of \textit{ad\_title, ad\_category, search\_query, search\_id} and other features. The \textit{search\_id} is the same as \textit{page\_id}, thus we consider that ads with same \textit{search\_id} are under the same page to construct the user's historical page sequence. The dataset has been widely used as benchmarks for CTR Prediction. Following other works ~\cite{ouyang2019deep, ouyang2019representation}, we use the ad logs from 2015-04-28 to 2015-05-18 for training, those on 2015-05-19 for validation, and those on 2015-05-20 for testing. Moreover, only users with more than one page are kept.

\item \textbf{Taobao dataset}: Taobao dataset comes from search logs from 2020-11-13 to 2020-11-27 in Taobao App. We hold the first 14 days as the training dataset and leave the last day for testing. The clicked items are kept as the positive samples, and the unclicked items are downsampled as the negative samples. For the historical page sequence, we select past search result pages which are in the  same product category (e.g., clothes, Electronic, etc.) with user's current input query. For each sample, we will use the user's recent ten pages as the page sequence.

\end{itemize}

\subsection{\textbf{Baseline Methods.}} We compare our RACP model to six baselines in different aspects. 

The first two baselines model the feature interaction through neural CF (collaborative filtering) and FM (factorization machine), they do not take user historical behaviors into considerations:

\begin{itemize}
    \item \textbf{NCF~\cite{he2017ncf}}: Neural Collaborative Filteringis a typical neural collaborative filtering approach which models the feature interaction through neural networks. 
    \item \textbf{DeepFM~\cite{guo2017deepfm}}: DeepFM  combines factorization machines and deep learning for CTR prediction.
\end{itemize}

Another two baselines incorporate user's historical clicked behaviors:
\begin{itemize}
    \item \textbf{YoutubeNet~\cite{covington2016deep} }: YoutubeNet is a model that uses users' watched videos sequence for video recommendation. It treats users' historical behaviors equally and utilizes the average pooling operation. 
    \item \textbf{DIN~\cite{zhou2018deep}}: DIN is a classical model for session-based recommendation. It considers the weights of items in user's historical behaviors with attention mechanism.
\end{itemize}

The last two methods consider both the user's historical clicked behaviors and unclicked behaviors:
\begin{itemize}
    \item \textbf{DFN~\cite{xie2020deep}}: DFN treats click behaviors as strong feedback to guide the positive preference extraction from unclicked behavior sequence. 
    \item \textbf{DSTN~\cite{ouyang2019deep}}: DSTN  considers the clicked and unclicked behaviors as heterogeneous auxiliary data to help the user preference modeling.
\end{itemize}

As for other work about page context~\cite{zhuang2018globally, ai2018learning, pei2019personalized} referred to in Section \ref{label:reated_pagecontext}, they are used to the re-ranking task. Neither of them focuses on CTR prediction. So they cannot be used as baseline methods.

\subsection{\textbf{Implementation Details.}} 
We adopt Adam~\cite{kingma2015adam} as the optimizer. 
We use the Dropout strategy \cite{srivastava2014dropout} to alleviate the overfitting issue in optimizing deep neural network models.

For the Avito dataset, we use PyTorch to implement our model and deploy it on Tesla V100-PCIE GPU with 16G memory. The hyper-parameter settings are as following: batch size $B=2048$, embedding dimension $D=10$, number of the sequence of pages $T=5$, number of items per page $L_a=5$. The learning rate is set as $\eta=0.001$. The Leaky ReLU slope is 0.2. In deep neural network layer, we adopt two layer neural networks and the hidden units are $L_1 = 200, L_2 = 80 $. 

For the Taobao dataset, the hyper-parameter settings are as following: batch size $B=1024$, number of the sequence of pages $T=10$, number of items per page $L_a=14$. The learning rates is set as $\eta=0.01$. The Leaky ReLU slope is 0.2. The output size of $s$ and the hidden size of attention module are set as 128. And we train the RACP on Taobao dataset in a large-scale distributed machine learning platform.

\subsection{\textbf{Evaluation Metrics.}} As previous work did, we utilize a widely-adopted metric Area Under Curve(AUC)~\cite{zhou2018deep,zhou2019deep} for evaluation in the CTR prediction task. Furthermore, we adopt an variation of AUC named Page-View AUC (PV\_AUC), which measures the goodness of intra-page order by averaging AUC over pages and is shown to be more relevant to  perfomance in online search ranking system. In particular, this evaluation criterion is also more difficult so that the gap among different methods may be smaller.
We adopt this metric in our experiments on Taobao dataset. It is calculated as follows:
\begin{equation}
    PV\_AUC = \frac{\sum^{n_p}_{i=1} \text{AUC}_i}{ \sum^{n_p}_{i=1} \text{1}}
\end{equation}
where $n_p$ is the number of pages, $\text{AUC}_i$ is the intra-page AUC of $i$-th page.
Following \cite{yan2014coupled}, we further bring a RelaImpr to measure the relative improvements over baseline model. Since AUC is 0.5 from a random strategy, the RelaImpr in this task is formalized as:\\
\begin{equation}
    RelaImpr = \frac{AUC(measured model)-0.5}{AUC(base model)-0.5} - 1
\end{equation}


\section{EXPERIMENTAL RESULTS}

We conduct extensive experiments to answer the following research questions: \\
\textbf{RQ1}: How does RACP perform compared with state-of-the-art models for CTR Prediction? \\
\textbf{RQ2}: Are the key components in RACP necessary for improving performance? \\
\textbf{RQ3}: How do hyper-parameters in RACP impact the performance of CTR prediction ? \\
\textbf{RQ4}: How does RACP perform in the real-world e-commerce search engine system? \\
\textbf{RQ5}: Can RACP provide meaningful interpretation of the prediction results?

\subsection{Overall Performance: RQ1}

The main results of the performance of our method and baselines are summarized in Table~\ref{main_results}. RACP outperforms state-of-the-art methods across the two datasets. The experimental results verify the effectiveness and generality of our proposed method.
From this table, we can find that:

\begin{itemize}
\item The models NCF~\cite{he2017ncf} and DeepFM~\cite{guo2017deepfm} do not use any user's behavior sequence.
They only use the categorical and statistical profile features and focus on feature interaction with neural networks. It can be seen that these methods with no user sequence are inferior to other methods which exploit user's behavior sequence.
This verifies the importance of modeling user's historical behaviors.

\item The models YoutubeNet~\cite{covington2016deep} and DIN~\cite{zhou2018deep} are popular methods in industry which only consider the user's positive feedback. Youtube-Net, which simply averages the user's history behaviors, outperforms the NCF and DeepFM about 5\% in RelaImpr. 
DIN, which utilizes the attention mechanism over the click sequence, obtains slightly better results. 
These two methods outperform NCF and DeepFM by incorporating the click sequence. Compared with the models with both click and unclick behavior, their performance is lower. The results show that the click sequence is essential for performance but is still insufficient because they omit the user's complete interaction history.

\item The methods DFN~\cite{xie2020deep} and DSTN~\cite{ouyang2019deep} are the previous state-of-the-art methods for modeling both the positive and negative feedback. They outperform YoutubeNet and DIN by modeling the click and unclick sequence separately. However, RACP obtains better results compared with DSTN and DFN by organizing the user's complete action as page sequence and modeling the intra-page page-context information  and the inter-page interest evolution. This verifies the success of incorporating the Contextualized Page-wise feedback information and our model.

\item 
In the two datasets, RACP achieves the best result in different metrics. The relative improvement of RACP over NCF is 8.92\% and 14.50\% on Avito and Taobao datasets respectively. Even compared with the best baseline in the table, RACP still obtains about 1\% gains of the absolute AUC score. 
This is a significant improvement for industrial applications where 0.1\% absolute AUC gain is remarkable~\cite{zhou2018deep,zhou2019deep}.

\end{itemize}

\subsection{Ablation Study: RQ2}
To investigate the effectiveness of components in our method RACP, we conduct extensive ablation studies in the large-scale Taobao dataset and report the results in Table ~\ref{page context ablation}.


\subsubsection{\textbf{Intra-page context}}
To validate the influence of the context information between clicked items and unclicked items in one page, we investigate the performance by removing different types of information. 

In the ``w/o action type '' variant, we put the clicked items and unclicked items together but remove the action types (click or unclick) information.
The user's behaviors are simply organized as an exposure sequence. 
The AUC drops significantly to 0.7585. 
This shows that just putting the exposed items together is not enough. 
The improvements come from the mutual influence between the items of different types.
This demonstrates the importance of the action types in the feedback.

In the ``w/o unclicked items'' variant, the model only uses clicked items in the page-sequence. 
Its AUC score is lower than RACP by 0.6\% points. 
The ``w/o clicked items" variant, which only use the unclicked items, is also inferior to RACP. 
This shows that it is necessary to include the user's complete interaction history with different types of actions.

In the ``split clicked and unclicked items into two sequences'' variant, 
We split the click and unclick behaviors into two individual sequences. The performance is still lower than RACP, which indicates the effectiveness of modeling the page-contextualized influence between clicked items and unclicked items in one page. 



\begin{table}[tbp]
\caption{Ablation study of Intra-page Context.}
\begin{tabular}{lll}
\toprule
Models                 & AUC      \\ \midrule
RACP                        & \textbf{0.7623}   \\ \midrule
w/o action type         & 0.7585   \\ 
w/o unclicked items         & 0.7558   \\
w/o clicked items           & 0.7577   \\
split clicked and unclicked items into two sequences & 0.7609   \\ 
\bottomrule
\end{tabular}
\label{page context ablation}
\end{table}

\subsubsection{\textbf{Model structure design.}}
We also conduct the ablation experiments on the structure of the model design and report the results in Table ~\ref{structure comarison}.


In the ``w/o inter-page Interest backtracking layer'' variant, we remove the interest backtracking layer, and simply uses $Q_{T}$  for all historical pages. 
Its performance (0.7613) is inferior to RACP because of the loss of modeling the interests consistency between two pages.
This validates the superiority of the interest backtracking layer. 
In the `Transformer + w/o inter-page Interest backtracking layer'' variant, we remove the interest backtracking layer as the ``w/o inter-page Interest backtracking layer'' variant, and use Transformer~\cite{vaswani2017attention,chen2019behavior,sun2019bert4rec,gu2020deep} in the Intra-page Context-aware Interest Layer and find that it has no significant improvement in our experiments.
This can be explained as the context information of the page is more related to comparison than combination. 
In the ``HGRU on page\_seq'' variant, we replace the attention with HGRU~\cite{quadrana2017personalizing} to model the hierarchical page sequence. This method use the GRU to capture the interest evolution similar as the DIEN~\cite{zhou2019deep}. The experiment results reveal that the variant's performance is worse than RACP. 

In the ``flatten the page seq + one-layer attention" variant, we flatten all the items in the  sequence of pages, treat them as a long sequence, and then use vanilla attention to aggregate the flattened sequence. 
We find that its performance is lower than RACP in AUC about 0.1\%, which shows the effectiveness of modeling the hierarchical structure of page sequence with two-level hierarchical attention module. 
In the ``mean\_pooling on inter-page aggregation layer'' variant, we exploit mean pooling in inter-page interest aggregation layer. The result demonstrates the effectiveness of Inter-page aggregation layer.


From these ablation studies, we verify the effectiveness of components in our method RACP.

\begin{table}[tbp]
\caption{Ablation study of model design.}
\begin{tabular}{lll}
\toprule
Models         & AUC     \\ \hline
RACP                & \textbf{0.7623}  \\ \midrule
w/o inter-page Interest backtracking layer      & 0.7613 \\  
Transformer + w/o inter-page Interest backtracking layer     & 0.7613  \\ 
HGRU on page\_seq   & 0.7580  \\ 
flatten the page seq + one-layer attention       & 0.7611  \\ 
mean\_pooling on inter-page aggregation layer & 0.7598  \\
\bottomrule
\end{tabular}
\label{structure comarison}
\end{table}

\subsection{Parameter Sensitivity: RQ3}
To study whether the performance of RACP is consistent under different settings and hyper-parameters, we conduct parameter sensitivity analysis.

\begin{figure}[tbp]
    \centering
    \includegraphics[width=0.9\linewidth]{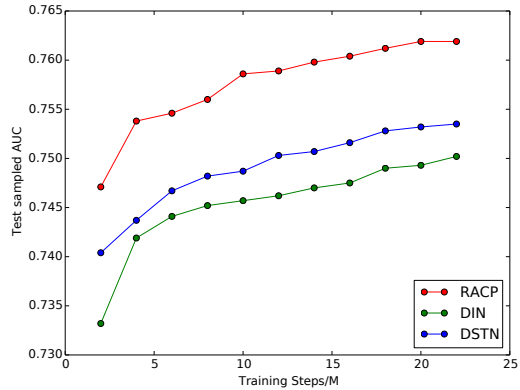}
    \caption{The improvement of RACP over baselines in different training steps on Taobao dataset.}
    \label{fig:training steps}
\end{figure}

In Figure~\ref{fig:training steps}, we compare the performance of RACP with baselines in the different training steps. We can see that RACP outperforms the DIN and DSTN consistently across the whole training process. 

\begin{figure}[tbp]
    \centering
    \includegraphics[width=0.9\linewidth]{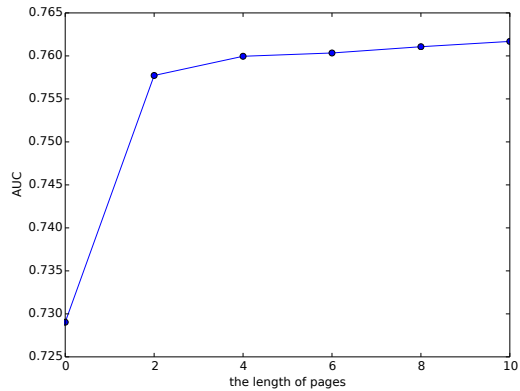}
    \caption{Performance of RACP with different length of pages on Taobao dataset.}
    \label{fig:page_len}
\end{figure}

Figure~\ref{fig:page_len} shows the results of RACP with different page length, which is the numbers of SRPs in the sequence. 
We can find that: The usage of page sequence can improve the AUC. As the page length increase, the performance will be improved.

\begin{figure}[tbp]
    \centering
    \includegraphics[width=0.95\linewidth]{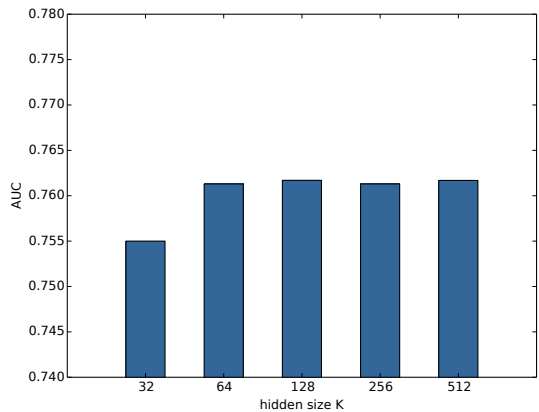}
    \caption{Parameter sensitivity of hidden size K on Taobao dataset.}
    \label{fig:hidden_size}
\end{figure}

Figure ~\ref{fig:hidden_size} illustrated the performance of RACP under different hidden size $K$ of the attention module, which is the same with the size of user's final interest vector $s$.
As the hidden size increases from 32 to 128, the AUC increases continually.
When the hidden size increases from 128 to 512, the AUC is relatively stable.

\subsection{Online A/B Testing : RQ4}

We conduct online A/B testing in Taobao Search for 7-days.
Compared with the strongest deplolyed online baseline, RACP significantly~(p-value < 0.01) improves the order count and Gross Merchandise Value (GMV) by +0.66\% and +0.9\% respectively (the noise level is less then 0.1\% according to the online A/A test).
RACP has been deployed online successfully in Taobao Search.

\begin{figure}[tbp]
     \centering
     \includegraphics[width=0.85\linewidth]{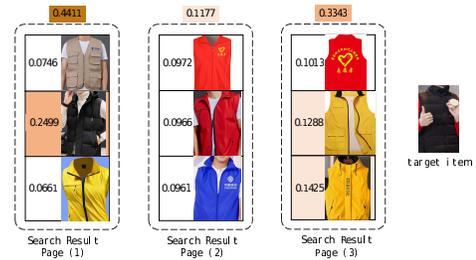}
     \caption{Case study of RACP on Taobao dataset. The darker the color is, the greater the attention weight is.}
     \label{fig:case}

\end{figure}

\subsection{Case Study: RQ5}
In this section, we investigate whether RACP can provide meaningful interpretation of the prediction using case study. 

In Figure~\ref{fig:case}, we plot the intra-page and inter-page attention weights of the top 3 items in the recent 3 pages, and find that:  If the items in a SRP page are more related to the target item, the weight of the SRP page will be higher. In each SRP page, the attention weight of an item in the page is higher if the item is more similar to the target item.
This demonstrates that RACP can accurately model users’ interest from Contextualized Page-wise Feedback, and our method has good interpretation ability for the prediction results.

\section{Conclusion}
In this paper, we propose the idea of modeling Users’ Contextualized Page-wise Feedback for CTR prediction in e-commerce search ranking.
We proposed to organize the user's behavior as a sequence of page-wise feedback and designed a RACP model to capture the user's  interest. RACP extracts the user's interest in each page using the Intra-page Context-aware Interest Layer, and captures user's dynamic interest in the session using the Inter-page Interest Backtracking Layer, and aggregates the user's final interest vector using a Page-level Interest Aggregation Layer.
We conduct extensive experiments on both public dataset and production dataset. 
The experimental results validate the effectiveness of our proposed model. 

\bibliographystyle{ACM-Reference-Format}
\balance
\bibliography{reference}

\appendix

\end{document}